\documentclass[preprintnumbers, prd, showpacs, floatfix,twocolumn,
preprintnumbers, letterpaper, superscriptaddress,nofootinbib]{revtex4}
\usepackage{graphicx,epsfig}
\usepackage{amsmath}
\usepackage{amssymb}
\usepackage{subfigure}

\begin{document}

\title{Higher-dimensional evolving wormholes satisfying the null energy
condition}
\author{ Mahdi Kord Zangeneh}
\email{mkzangeneh@shirazu.ac.ir}
\affiliation{Physics Department and Biruni Observatory, Shiraz University, Shiraz 71454,
Iran}
\author{Francisco S. N. Lobo}
\email{flobo@cii.fc.ul.pt}
\affiliation{Centro de Astronomia e Astrof\'{\i}sica da Universidade de Lisboa, Campo
Grande, Edific\'{\i}o C8, 1749-016 Lisboa, Portugal}
\author{Nematollah Riazi}
\email{n_riazi@sbu.ac.ir}
\affiliation{Physics Department, Shahid Beheshti University, Evin, Tehran 19839, Iran}
\date{\today }

\begin{abstract}

In this work, we consider the possibility of expanding wormholes in higher-dimensions, which is an important ingredient of modern theories of fundamental physics. An important motivation is that non-trivial topological objects such as microscopic wormholes may have been enlarged to macroscopic sizes in an expanding inflationary cosmological background. Since the Ricci scalar is only a function of time in standard cosmological models, we use this property as a simplifying assumption.  More specifically, we consider a particular class of wormhole solutions corresponding to the choice of a spatially homogeneous Ricci scalar. The possibility of obtaining solutions with normal and exotic matter is explored and we find a variety of solutions including those in four dimensions that satisfy the null energy condition (NEC) in specific time intervals. In particular, for five dimensions, we find solutions that satisfy the NEC throughout the respective evolution.

\end{abstract}

\maketitle

%%%%%%%%%%%%%%%%%%%%%%%%%%%%%%%%%%%%%%%%%%%%%%%%%%%%%%%%%%%%%%%%%%%%%%%%%%%%%%%%%%%%

\section{Introduction}

%%%%%%%%%%%%%%%%%%%%%%%%%%%%%%%%%%%%%%%%%%%%%%%%%%%%%%%%%%%%%%%%%%%%%%%%%%%%%%%%%%%%

Wormhole physics dates back to the formulation of General Relativity (GR).
Indeed, in 1916, months after Einstein presented his gravitational field
equations, Karl Schwarzschild found the first solution of Einstein's
equations, which described the gravitational field of a vacuum non-rotating
spherically symmetric solution. In the same year, Ludwig Flamm published a paper in which the geometry of the 
Schwarzschild solution was studied more closely. He pointed out a ``tunnel-shaped'' nature of space near the 
Schwarzschild radius, this being perhaps the first move towards the modern concept of the ``throat'' in 
wormholes \cite{reff}.

Paging through the literature, one finds next that tunnel-like solutions
were considered, in 1935, by Einstein and Rosen, where they constructed an
elementary particle model represented by a \textquotedblleft
bridge\textquotedblright\ connecting two identical sheets \cite{1}. They
considered the possibility that fundamental particles such as the electron
could be represented as microscopic spacetime tunnels that convey fluxes of
the electric field. These tunnels were later denoted the \textit{%
Einstein-Rosen bridge}. In fact, the Einstein-Rosen bridge is a coordinate
artifact arising from choosing a coordinate patch, which is defined to
double-cover the asymptotically flat region exterior to the black hole event
horizon.

The field had lain dormant for about twenty years when, in 1955, John Wheeler, who
was beginning to be interested in topological issues in GR, explored
solutions of the coupled Einstein--Maxwell equations, which he denoted
\textit{geons} (gravitational-electromagnetic entities) \cite{2}. These were
considered to be objects of the quantum foam connecting different regions of
spacetime at the Planck scale. However, the term `wormhole' was only used
for the first time in 1957 \cite{Misner:1957mt}, where Misner and Wheeler
presented a tour de force wherein Riemannian geometry of manifolds of
nontrivial topology was investigated with an ambitious view to explaining
all of physics. The aim was to use the source-free Maxwell
equations, coupled to Einstein gravity, with nontrivial topology, to build
models for classical electrical charges and all other particle--like
entities in classical physics.

Subsequently to the geon concept, several wormhole solutions were obtained
and discussed within different contexts \cite{6288}. However, it was only in
1988 that the full-fledged renaissance of wormhole physics took place
through the seminal Morris-Thorne paper \cite{4}, and the theme is still in
full flight. Morris and Thorne, considered static and spherically symmetric
traversable wormholes, and thoroughly analysed their fundamental properties.
It was found that these traversable wormholes possess a stress-energy tensor
that violates the null energy condition (NEC), a property that was denoted
\textit{exotic matter}. Beside being hypothetical short-cuts in spacetime
and consequently useful for inter- and intra-universe travel, they were
found to possess other intriguing applications, such as, the usage for
time-travel \cite{9a} and investigating the interior of a black hole \cite{9}%
, amongst others.

Thus, a fundamental ingredient for the Morris-Thorne wormhole, i.e., for
static and spherically symmetric wormhole solutions, is the violation of the
NEC \cite{4,5}. Exotic matter is particularly troublesome for measurements
made by observers traversing through the throat with a radial velocity close
to the speed of light, as for sufficiently high velocities, $v \rightarrow c$%
, the observer will measure a negative energy density \cite{4}. Although
classical forms of matter are believed to obey these energy conditions, it
is a well-known fact that they are violated by certain quantum fields, such
as the Casimir effect. In fact, the recent discovery that the universe is
undergoing an accelerated expansion \cite{dark} may be due to an exotic
cosmic fluid that lies in the phantom regime. The realization of this fact
has led to the study of wormhole solutions supported by different kinds of
phantom fluids (see, for instance \cite{phantomWH}). Indeed the violation of
the energy conditions is a subtle issue, as almost all known and physically
possible forms of matter satisfy these energy conditions, and we recall that
their imposition is one of the necessary assumptions for proving the
Hawking-Penrose singularity theorems \cite{15,17}.

Thus, one may adopt the approach of minimizing the usage of exotic matter
\cite{Visser:2003yf,normal}. In this context, a plethora of solutions have
been investigated, using a wide variety of approaches \cite%
{diffpo,diffpo0,diffpo2,diffpo3,18}. More specifically, in rotating
solutions it was found that the exotic matter lies in specific regions
around the throat, so that it is possible for a certain class of infalling
observers to move around the throat as to avoid the exotic matter supporting
the wormhole \cite{Teo:1998dp}. Using the thin shell formalism, solutions
where the exotic matter is concentrated at the throat have also been
extensively investigated \cite{thinshell}. In the context of modified
gravity it was shown that one may impose that the matter threading the
wormhole satisfies the energy conditions, so that it is the higher order
curvature terms that sustain these exotic geometries \cite{modgravWH}.
Astrophysical signatures have also been explored in the literature \cite%
{Harko:2008vy}.

For dynamic wormholes, the NEC, or more precisely the averaged null energy
condition, can be avoided in certain regions \cite%
{hochvisserPRL98,hochvisserPRD98,kar,kar-sahdev,Kim-evolvWH,Arellano:2006ex}%
. A particularly interesting case is that of a wormhole in a time-dependent
inflationary background \cite{romanLambda}, in which the primary goal was to
use inflation to enlarge an initially small and possibly submicroscopic
wormhole. It is also possible that the wormhole will continue to be enlarged
by the subsequent FRW phase of expansion. One could perform a similar
analysis to \cite{romanLambda} by replacing the deSitter scale factor by an
FRW scale factor \cite{kar,kar-sahdev,Kim-evolvWH}. In particular, in \cite%
{kar,kar-sahdev} specific examples for evolving wormholes that exist only
for a finite time were considered, and a special class of scale factors
that exhibit `flashes' of the WEC violation were also analyzed.

The present paper investigates the possibility and naturalness of expanding
wormholes in higher dimensions which is an important ingredient of the
modern theories of fundamental physics, such as string theory, supergravity,
Kaluza-Klein, and others. One of our motivations for considering wormhole solutions
in an expanding cosmological background refers to the inflation theory \cite%
{guth} where the quantum fluctuations in the inflaton field are considered
as the seed of large scale structures in the universe. As mentioned above,
the non-trivial topological objects such as microscopic wormholes may have
been formed during inflation and enlarged to macroscopic ones as the
universe expanded \cite{romanLambda}. We also explore the possibility that
these higher-dimensional wormholes satisfy the NEC, and we explicitly show
that this is indeed the case.

This paper is organized in the following way: In Section \ref{secII}, we
present the $(n+1)-$dimensional field equations for the specific case of a
spatially-independent curvature scalar. In Section \ref{secIII}, we analyse
the two-way traversability conditions of the wormhole structure. In Section %
\ref{secIV}, we explore wormhole solutions in different expansionary
regimes, and finally in Section \ref{conclusion}, we conclude.

%%%%%%%%%%%%%%%%%%%%%%%%%%%%%%%%%%%%%%%%%%%%%%%%%%%%%%%%%%%%%%%%%%%%%%%%%%%%%%%%%%%%

\section{Action, Field Equations and $(n+1)-$Dimensional solutions}

\label{secII}
%%%%%%%%%%%%%%%%%%%%%%%%%%%%%%%%%%%%%%%%%%%%%%%%%%%%%%%%%%%%%%%%%%%%%%%%%%%%%%%%%%%%

The action of GR in $(n+1)$--dimensions is written as
\begin{equation}
S=\int {\mathrm{d}}^{n+1}x\sqrt{-g}\left(\frac{1}{2}\mathcal{R}+\mathcal{L}%
_{m} \right),  \label{act1}
\end{equation}%
where $\mathcal{R}$ is the scalar curvature and $\mathcal{L}_{m}$ is the
matter Lagrangian density; we have considered $c=8\pi G=1$. Varying this
action with respect to the metric, we obtain the $(n+1)$-dimensional
Einstein equations $G_{AB}=T_{AB}$, where $(A,B=0 ... n)$, and $T_{AB}$ is
the matter stress-energy tensor.

Since we are looking for expanding wormhole solutions in a cosmological
background, we use the metric%
\begin{equation}
ds^{2}=-dt^{2}+R(t)^{2}\left[ \frac{dr^{2}}{1-a(r)}+r^{2}d\Omega _{n-1}^{2}%
\right] ,  \label{metn}
\end{equation}%
in which $R(t)$ is the scale factor and $a(r)$ is an unknown dimensionless
function, defined as $a(r)=b(r)/r$, where $b(r)$ denotes the shape function
\cite{4}. Note that this metric is a generalization of the
Friedmann-Robertson-Walker (FRW) metric, although being less symmetric than
the latter. With this generalization, metric (\ref{metn}) is still isotropic about
the center of the symmetry, though not necessarily homogeneous. When the
dimensionless shape function vanishes, $a(r)\rightarrow 0$ the metric (\ref%
{metn}) reduces to the flat FRW metric; and as $R(t)\rightarrow \mathrm{const%
}$ it approaches the static wormhole metric.

To see that the ``wormhole'' form of the metric is preserved with time,
consider an embedding of $t=\mathrm{const}$ and $\theta_{(n-2)}=\pi/2$
slices of the spacetime given by Eq. (\ref{metn}), in a flat 3-dimensional
Euclidean space with metric
\begin{equation}
ds^2=d{\bar{z}}^2+d{\bar{r}}^2+{\bar{r}}^2\,{d\phi}^2\,.  \label{barredslice}
\end{equation}
In this context, the metric of the wormhole slice is
\begin{equation}  \label{slice}
ds^2={\frac{R^2(t)\,{dr^2}}{{1-a(r)}}} + R^2(t)\, r^2\,d\phi^2\,.
\end{equation}
Now, comparing the coefficients of ${d\phi}^2$, one has
\begin{eqnarray}
\bar{r}&=&{R(t)\,r}\big|_{t=\mathrm{const}} \,,  \label{coef1:phi} \\
{d\bar{r}}^2&=&R^2(t)\,{dr}^2\big|_{t=\mathrm{const}} \,.  \label{coef2:phi}
\end{eqnarray}
It is important to keep in mind, in particular, when considering
derivatives, that Eqs. (\ref{coef1:phi})-(\ref{coef2:phi}) do not represent
a ``coordinate transformation'', but rather a ``rescaling'' of the $r$
coordinate on each $t=\mathrm{constant}$ slice \cite{romanLambda}.

With respect to the ${\bar{z}},{\bar{r}},\phi$ coordinates, the ``wormhole''
form of the metric will be preserved if the metric on the embedded slice has
the form
\begin{equation}  \label{WHslice}
ds^2={\frac{{d{\bar{r}}^2}}{{1-{\bar{a}(\bar{r})}}}} + {\bar{r}}^2{d\phi}%
^2\,,
\end{equation}
where $\bar{a}(\bar{r}_0)=1$, i.e., $\bar{b}(\bar{r})$ has a minimum at some
$\bar{b}(\bar{r}_0)=\bar{r}_0$. Equation (\ref{slice}) can be rewritten in
the form of Eq. (\ref{WHslice}) by using Eqs. (\ref{coef1:phi})-(\ref%
{coef2:phi}) and
\begin{equation}
\bar{a}(\bar{r})=R(t)\,a(r).  \label{bar:b}
\end{equation}
The evolving wormhole will have the same overall size and shape relative to
the ${\bar{z}},{\bar{r}},\phi$ coordinate system, as the initial wormhole
had relative to the initial $z,r,\phi$ embedding space coordinate system.
This is due to the fact that the embedding space corresponds to $z,r$
coordinates that ``scale'' with time (each embedding space corresponds to a
particular value of $t=\mathrm{constant}$). Following the embedding
procedure \cite{4}, using Eqs. (\ref{barredslice}) and (\ref{WHslice}), one
deduces that
\begin{equation}
{\frac{{d{\bar{z}}}}{{d{\bar{r}}}}} =\pm\left({\frac{1}{{\bar{a}(\bar{r})}}}%
-1\right)^{-1/2} ={\frac{{dz}}{{dr}}} \,.  \label{barredembedding}
\end{equation}
which implies
\begin{eqnarray}
\bar{z}(\bar{r})=\pm R(t)\,z(r)\,.  \label{embed:relation}
\end{eqnarray}

Therefore, we see that the relation between the embedding space at any time $%
t$ and the initial embedding space at $t=0$, from Eqs. (\ref{coef2:phi}) and
(\ref{embed:relation}), is given by the following
\begin{equation}
ds^2=d{\bar{z}}^2+d{\bar{r}}^2+{\bar{r}}^2\,{d\phi}^2 =R^2(t)\,[dz^2+dr^2+r^2%
{d\phi}^2]\,.
\end{equation}
Relative to the ${\bar{z}},{\bar{r}},\phi$ coordinate system the wormhole
will always remain the same size, as the scaling of the embedding space
compensates for the evolution of the wormhole. However, the wormhole will
change size relative to the initial $t=0$ embedding space.

Writing the analog of the ``flaring out condition'' \cite{4} for the
evolving wormhole we have $d\,^2{\bar{r}(\bar{z})}/d{\bar{z}}^2>0$, at or
near the throat. From Eqs. (\ref{coef1:phi}), (\ref{coef2:phi}), (\ref{bar:b}%
), and (\ref{barredembedding}), it follows that
\begin{equation}
{\frac{{d\,^2{\bar{r}(\bar{z})}}}{{d{\bar{z}}^2}}} =\frac{1}{R(t)}\,\left(-
\frac{a^{\prime }}{2a^2} \right) =\frac{1}{R(t)}\,{\frac{{d\,^2r(z)}}{{dz^2}}%
}>0\,,  \label{barred:flareout}
\end{equation}
at or near the throat, where the prime denotes the derivative with respect
to $r$. Note that this also implies $a^{\prime }<0$, at or near the throat.
Taking into account Eqs. (\ref{coef1:phi}), (\ref{bar:b}), and ${\bar{b}}%
^{\prime }(\bar{r})={d\bar{b}}/{d\bar{r}}=b^{\prime }(r)=db/dr$, one may
rewrite the right-hand-side of Eq. (\ref{barred:flareout}) relative to the
barred coordinates as
\begin{equation}
{\frac{{d\,^2{\bar{r}(\bar{z})}}}{{d{\bar{z}}^2}}} =-{\frac{\bar{a}^{\prime }%
}{{2{\bar{a}}^2}}}>0\,,  \label{barred:flareout2}
\end{equation}
or $\bar{a}<0$ at or near the throat. One verifies that using the barred
coordinates, the flaring out condition Eq. (\ref{barred:flareout2}), has the
same form as for the static wormhole.

Thus, it can be shown that metric (\ref{metn}) represents a traversable
wormhole provided%
\begin{eqnarray}
a(r_{0}) =1, \qquad a(r) <1, \qquad a^{\prime }(r) <0,  \label{3-3}
\end{eqnarray}%
where $r_0$ is the wormhole throat, which represents a minimum radius in the
wormhole space-time \cite{4}. The second condition is imposed in order to
avoid a change in the metric signature. The third condition is the
flaring-out condition and plays a fundamental role in the analysis of the
violation of the energy conditions.

Note that the comoving radial distance defined by
\begin{equation}
l(r)=\pm \int\limits_{r_{0}}^{r}{\frac{\mathit{\ }{\mathrm{d}}r}{\sqrt{1-a{%
(r)}}}}  , \label{4-1}
\end{equation}%
should be real and finite everywhere in spite of the fact that the $rr$%
-component of the covariant metric diverges at the throat; the $\pm $ signs
denote the upper and lower parts of the wormhole.

The energy-momentum tensor is $T_{B}^{A}={\mathrm{diag(}}-\rho ,$ $P_{r},$ $%
P_{t},$ $P_{t},$ $...{\mathrm{)}}$, so that using the Einstein field
equations and Eq. (\ref{metn}), the ($n+1$)--dimensional field equations are
satisfied by the following stress-energy profile
\begin{eqnarray}
\rho (r,t) &=&{\frac{\left( n-1\right) \left( n-2\right) a\left( r\right) }{%
2R\left( t\right) ^{2}{r}^{2}}}+{\frac{\left( n-1\right) a^{\prime }(r)}{%
2R\left( t\right) ^{2}r}}  \notag \\
&&+\,{\frac{n\left( n-1\right) \dot{R}\left( t\right) ^{2}}{2R\left(
t\right) ^{2}},} \\
P_{r}(r,t) &=&-{\frac{\left( n-1\right) \ddot{R}\left( t\right) }{R\left(
t\right) }}-{\frac{\left( n-1\right) \left( n-2\right) \dot{R}\left(
t\right) ^{2}}{2R\left( t\right) ^{2}}}  \notag \\
&&-{\frac{\left( n-1\right) \left( n-2\right) a\left( r\right) }{2R\left(
t\right) ^{2}{r}^{2}},} \\
P_{t}(r,t) &=&-{\frac{\left( n-2\right) a^{\prime }(r)}{2R\left( t\right)
^{2}r}}-{\frac{\left( n-1\right) \left( n-2\right) \dot{R}\left( t\right)
^{2}}{2R\left( t\right) ^{2}}}  \notag \\
&&-{\frac{\left( n-1\right) \ddot{R}\left( t\right) }{R\left( t\right) }}-{%
\frac{\left( n-2\right) \left( n-3\right) a\left( r\right) }{2R\left(
t\right) ^{2}{r}^{2}},}
\end{eqnarray}%
where the overdot denotes a derivative with respect to time.

The Ricci scalar will play a fundamental role in our analysis, so we write
it down explicitly as
\begin{eqnarray}
\mathcal{R}&=&{\frac{2n\ddot{R}\left( t\right) }{R\left( t\right) }}+{\frac{%
\left( n-1\right) a^{\prime }(r)}{R\left( t\right) ^{2}r}}+{\frac{n\left(
n-1\right) \dot{R}\left( t\right) ^{2}}{R\left( t\right) ^{2}}}  \notag \\
&&+{\frac{\left( n-1\right) \left( n-2\right) a\left( r\right) }{R\left(
t\right) ^{2}{r}^{2}}.}  \label{ricci}
\end{eqnarray}%
Since the Ricci scalar is only a function of time in standard cosmological
models, it provides a motivation to use this property as a simplifying
assumption in our calculations, in the presence of a wormhole. In other
words, we are looking for classes of solutions corresponding to the choice
of a homogeneous Ricci scalar, i.e., $\partial \mathcal{R}/\partial r=0$,
which implies
\begin{equation}
{r}^{2}a^{\prime \prime }(r)+\left( n-3\right) ra^{\prime
}(r)-2(n-2)\,a\left( r\right) =0.
\end{equation}

The above differential equation yields the following solution
\begin{equation}
a(r)=\frac{r_{0}^{n-2}-kr_{0}^{n}}{r^{n-2}}+kr^{2},  \label{3}
\end{equation}%
where the condition $a(r_{0})=1$ was used to eliminate the integration
constant. We point out that although $k$ can, in principle, be a
continuous variable, we have used the fact that the space-time is
asymptotically FRW and applied the normalization $k=0,\pm 1$ for the
curvature constant. It is worthwhile to mention that it is common to
consider static wormholes supported by radiation that have a traceless
stress-energy tensor \cite{vanishing Ricci scalar}. In such a case, the
Ricci scalar vanishes if there is no cosmological constant within the
framework of GR. Our assumption leads to the same situation, if the scale
factor is assumed to be independent of time, i.e., for a static case.

As mentioned before, the dimensionless shape function $a(r)$ should satisfy
the conditions (\ref{3-3}). It is easy to show that for $k=0$ and $-1$ these
conditions are satisfied whereas for $k=+1$ they are not. Therefore we
continue our discussions using $k=0$ and $-1$ which present flat and open
universes, respectively.

With $a(r)$ in hand, given by Eq. (\ref{3}), one can rewrite the field
equations for the spatially flat background ($k=0$) as:
\begin{eqnarray}
\rho  &=&\rho _{(fb)},  \label{rhoflat} \\
P_{r} &=&-\frac{{\left( n-1\right) \left( n-2\right) \mathit{r}_{0}^{n-2}}}{2%
{{r}^{n}}R^{2}}+P_{(fb)},  \label{prflat} \\
P_{t} &=&\frac{\left( n-2\right) {\mathit{r}_{0}^{n-2}}}{2{r}^{n}R^{2}}%
+P_{(fb)},  \label{ptflat}
\end{eqnarray}%
where $\rho _{(fb)}$ and $P_{(fb)}$ are the respective \textquotedblleft
flat background\textquotedblright\ components given by
\begin{eqnarray}
\rho _{(fb)} &=&\,{\frac{n\left( n-1\right) \dot{R}^{2}}{2R^{2}}}, \\
P_{r(fb)} &=&P_{t(fb)}=P_{(fb)} \\
&=&-\frac{{\left( n-1\right) \,\ddot{R}}}{R}-\frac{{\left( n-1\right) \left(
n-2\right) \dot{R}^{2}}}{2R^{2}},  \label{bgflat}
\end{eqnarray}%
respectively.

For the specific case of the open background ($k=-1$), we have
\begin{eqnarray}
\rho &=&\rho _{(ob)},  \label{rhoopen} \\
P_{r}&=&-\frac{\left( n-1\right) \left( n-2\right) \left( {\mathit{r}}%
_{0}^{n}+{{\ \mathit{r}}}_{0}^{n-2}\right) }{2{r}^{n}R^{2}}+P_{(ob)}, \\
P_{t}&=&\,{\frac{\left( n-2\right) \left( {\mathit{r}}_{0}^{n}+{{\ \mathit{r}%
}}_{0}^{n-2}\right) }{2{r}^{n}R^{2}}+}P_{(ob)},
\end{eqnarray}%
where the $\rho _{(ob)}$ and $P_{(ob)}$ components correspond to the ``open
background'', and are given by
\begin{eqnarray}
\rho _{(ob)}&=&\rho _{(fb)}-\frac{n(n-1)}{2R^{2}}, \\
P_{(ob)}&=&P_{r(ob)}=P_{t(ob)}  \notag \\
&=&P_{(fb)}+\frac{\left( n-1\right) (n-2)}{2R^{2}}.  \label{bgopen}
\end{eqnarray}
Since our solutions are in a spherically symmetric cosmological background,
the components of the stress-energy tensor should be asymptotically
independent of $r$. It is easy to see this expected behaviour is obeyed by
them.

\section{Two-Way Traversability of Wormhole Structure}

\label{secIII}

One of the most interesting properties of a wormhole as pointed out by
Morris and Thorne \cite{9} is its two-way traversability. In this section,
some proofs will be presented to show that the wormholes discussed in this
paper are indeed two-way traversable.

%%%%%%%%%%%%%%%%%%%%%%%%%%%%%%%%%%%%%%%%%%%%%%%%%%%%%%%%%%%%%%%%%%%%%%%%%%%%%%%%%%%%

\subsection{Redshift of a co-moving source}

\label{sub-51}
%%%%%%%%%%%%%%%%%%%%%%%%%%%%%%%%%%%%%%%%%%%%%%%%%%%%%%%%%%%%%%%%%%%%%%%%%%%%%%%%%%%%

Consider a radially moving light signal emitted from a co-moving source. We
assume that the signal is emitted at $(t_{1},l_{1})$ ($l_{1}$ is a co-moving
coordinate) and received by a distant co-moving observer at $(t_{0},l_{0})$.
Using the metric (\ref{metn}) and Eq. (\ref{4-1}) for a radial beam, we have
\begin{equation}
\int_{t_{1}}^{t_{0}}\frac{dt}{R(t)}=\int_{l_{1}}^{l_{0}}dl,  \label{42}
\end{equation}%
where $l(r)$ is the comoving radial distance defined in Eq. (\ref{4-1}).
Note that $l_{0}$ and $l_{1}$ can belong to either side of the throat. It is
obvious that the rhs of Eq. (\ref{42}) is independent of time. Therefore,
the lhs should also be so and a signal which is emitted in an interval $\tau
_{1}$\ should be received in an interval $\tau _{0}$\ such that ($\tau
_{1},\tau _{0}\ll t_{1},t_{0}$)
\begin{equation}
\int_{t_{1}}^{t_{0}}\frac{dt}{R(t)}=\int_{t_{1}+\tau _{1}}^{t_{0}+\tau _{0}}%
\frac{dt}{R(t)}.
\end{equation}%
Since $\tau _{0}$\ and $\tau _{1}$\ are very short time intervals, one
deduces that
\begin{equation}
\frac{\tau _{0}}{\tau _{1}}=\frac{R(t_{0})}{R(t_{1})}=1+z,  \label{43}
\end{equation}%
where $R(t_{0})$ is the scale factor at the time of observation, $R(t_{1})$
is the scale factor at the time of emission, and $z$ is the cosmological
redshift. This shows that the redshift is the same as the cosmological
redshift and no extra redshift is caused by the wormhole. It remains to
examine whether the signal ever reaches the throat in a finite time or not.
This will be addressed below.

%%%%%%%%%%%%%%%%%%%%%%%%%%%%%%%%%%%%%%%%%%%%%%%%%%%%%%%%%%%%%%%%%%%%%%%%%%%%%%%%%%%%

\subsection{The behavior of radial geodesics}

\label{sub-52}
%%%%%%%%%%%%%%%%%%%%%%%%%%%%%%%%%%%%%%%%%%%%%%%%%%%%%%%%%%%%%%%%%%%%%%%%%%%%%%%%%%%%

Since $r$ is greater than $r_{0}$ on both sides of the throat, one cannot
trivially deduce whether the light signal passes through the throat or not
using the $r$\ coordinate. Therefore, we need to transform from the radial
coordinate $r$ to the comoving radial coordinate $l$ in order to analyse
this behavior. Using Eq. (\ref{4-1}), the analysis is more transparent.
Consider radial motion so that the geodesic equation reads
\begin{equation}
\frac{d^{2}l}{d\lambda ^{2}}+\frac{2}{R}\frac{dR}{dt}\frac{dt}{d\lambda }%
\frac{dl}{d\lambda }=0,  \label{49}
\end{equation}%
and
\begin{equation}
\frac{d^{2}t}{d\lambda ^{2}}+R\frac{dR}{dt}\left( \frac{dl}{d\lambda }%
\right) ^{2}=0.  \label{49-1}
\end{equation}%
Equation (\ref{49}) yields the first integral
\begin{equation}
\frac{dl}{d\lambda }=\frac{C}{R^{2}}.  \label{50}
\end{equation}%
Equation (\ref{50}) shows that $dl/d\lambda$ does not undergo a
sign change along the path and neither does it vanish. This shows that the
particle or signal continues its path, passes the throat and goes to the
other side of the wormhole and therefore the wormhole is two-way traversable.

%%%%%%%%%%%%%%%%%%%%%%%%%%%%%%%%%%%%%%%%%%%%%%%%%%%%%%%%%%%%%%%%%%%%%%%%%%%%%%%%%%%%

\subsection{Reachability of the wormhole throat}

%%%%%%%%%%%%%%%%%%%%%%%%%%%%%%%%%%%%%%%%%%%%%%%%%%%%%%%%%%%%%%%%%%%%%%%%%%%%%%%%%%%%

In order to prove that there is a finite proper distance between a specific
point and the throat, Eq. (\ref{4-1}) should be solved explicitly. For
arbitrary $n$, there is no analytical solution for the integral (\ref{4-1}).
Therefore, one could obtain $l(r)$ in the vicinity of the throat, which is
sufficient for our purpose. Using the approximate relation $1-\left(
r_{0}/r\right) ^{n-2}\approx (n-2)(r/r_{0}-1)$, one obtains
\begin{equation}
l(r)\approx \left\{
\begin{array}{l}
2\sqrt{\frac{r_{0}(r-r_{0})}{n-2}} \qquad \mathrm{for} \quad k=0 \\
2\,\sqrt{\frac{r_{0}(r-r_{0})}{\left( n-2+nr_{0}^{2}\right) }} \quad \mathrm{%
for} \quad k=-1%
\end{array}%
\right. ,
\end{equation}%
which are clearly finite distances. Therefore, the throat is not located at
spatial infinity and a finite time is required to reach it.

We should mention that horizons are theoretical constructs that
qualitatively have two main specific properties: first, they are one-way
membranes, and second, the corresponding redshift (as observed by a
distant observer) is infinite. More precise mathematical details can be
found in \cite{17,hor}. Since it is common to consider the singularities of
the metric as candidates of being horizons, one might ask whether the
coordinate singularity at the throat forms a horizon. Based on the
above-mentioned general qualitative features of horizons and according to
the results obtained in subsections (\ref{sub-51}) and (\ref{sub-52}), it is
justified that there is no horizon at or around the throat. The possibility
of the existence of a cosmological horizon, however, depends on the behavior
of the scale factor $a(t)$ and is not essentially affected by the presence
or absence of the wormhole.

%%%%%%%%%%%%%%%%%%%%%%%%%%%%%%%%%%%%%%%%%%%%%%%%%%%%%%%%%%%%%%%%%%%%%%%%%%%%%%%%%%%%

\section{Wormhole Solutions in Different Expansion Regimes}

\label{secIV}
%%%%%%%%%%%%%%%%%%%%%%%%%%%%%%%%%%%%%%%%%%%%%%%%%%%%%%%%%%%%%%%%%%%%%%%%%%%%%%%%%%%%

One of the properties of normal matter is that it satisfies the energy
conditions, in particular, the null energy condition (NEC) and the weak
energy condition (WEC). It was mentioned in the introduction that the matter
that supports the static wormhole geometry violates the NEC and is
therefore denoted `exotic matter'. The NEC requires that $T_{\mu \nu }k^{\mu
}k^{\nu }\geq 0$, where $k^{\mu }$ is \textit{any} null vector. In terms of
the energy density, radial pressure and tangential pressure the NEC becomes
\cite{0}
\begin{equation}
\rho +P_{r}\geq 0,\qquad \rho +P_{t}\geq 0.  \label{wec}
\end{equation}%
Note that the WEC, in addition to the conditions considered above, also
imposes a positive energy density, $\rho \geq 0$. In what follows, we
investigate the NEC for the wormhole solutions in different expansion
regimes.

%%%%%%%%%%%%%%%%%%%%%%%%%%%%%%%%%%%%%%%%%%%%%%%%%%%%%%%%%%%%%%%%%%%%%%%%%%%%%%%%%%%%

\subsection{Flat Background}

%%%%%%%%%%%%%%%%%%%%%%%%%%%%%%%%%%%%%%%%%%%%%%%%%%%%%%%%%%%%%%%%%%%%%%%%%%%%%%%%%%%%

In the case of a flat background, one can obtain two different solutions for
the scale factor $R(t)$ by applying the equation of state $P_{(fb)}=\omega
\rho _{(fb)}$, given by
\begin{equation}
R(t)= \left\{
\begin{array}{l}
A_{1}e^{\alpha t}, \quad \alpha >0 \quad \text{for} \quad \omega =-1 \\
A_{2}t^{\frac{2}{n(1+\omega )}}, \qquad \text{for} \quad \omega \neq -1%
\end{array}%
\right. .
\end{equation}%
The specific case of $\omega =-1$ represents the inflationary regime. The
case of $\omega \neq -1$ represents the radiation dominated and matter
dominated expansion regimes, by considering $\omega =1/3$ and $\omega =0$,
respectively.

Consider first the inflationary expansion regime, where by using Eqs. (\ref%
{rhoflat})-(\ref{bgflat}), one obtains
\begin{eqnarray}
\rho &=&\frac{\,n\left( n-1\right) {\alpha }^{2}}{2}, \\
\rho +P_{r}&=&-{\frac{\left( n-1\right) \left( n-2\right) {\mathit{r}}%
_{0}^{n-2}}{2{r}^{n}{A}_{1}^{2}}\mathrm{e}^{-2\,\alpha \,t},} \\
\rho +P_{t}&=&\,{\frac{\left( n-2\right) {\mathit{r}}_{0}^{n-2}}{2{r}^{n}{A}%
_{1}^{2}}\mathrm{e}^{-2\,\alpha \,t}.}
\end{eqnarray}%
It is clear that $\rho +P_{r}$ is always negative while $\rho $ and $\rho
+P_{t}$ are always positive. Therefore, the NEC is always violated. But $%
\rho +P_{r}$ tends to zero as $t$ increases and therefore the wormhole 
matter ranges from an exotic matter regime to normal matter over time.
\begin{figure}[tbp]
\centering
\includegraphics[width=7.0cm]{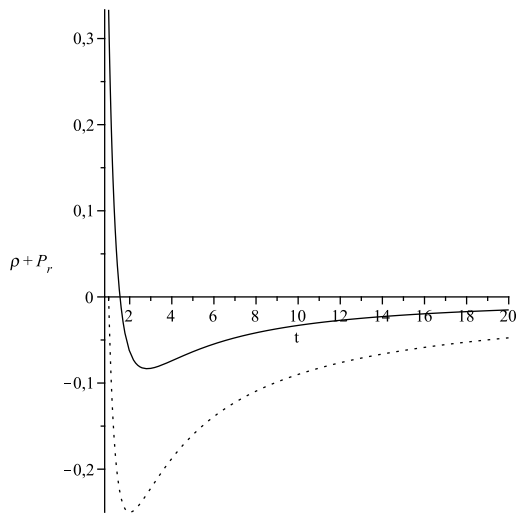}
\caption{For the flat background solution, the plot depicts the behavior of $%
\protect\rho +P_{r}$ at the throat, assuming $r_{0}=1$, with respect to
time, with $n=3$ and $A_{2}=1$ for $\protect\omega =0$ (solid curve) and $%
\protect\omega =1/3$ (dashed curve). The plot shows that although initially
there is normal matter at throat, the throat matter tends to an exotic
matter regime over time.}
\label{i1}
\end{figure}
\begin{figure*}[tbp]
\centering
\includegraphics[width=8cm]{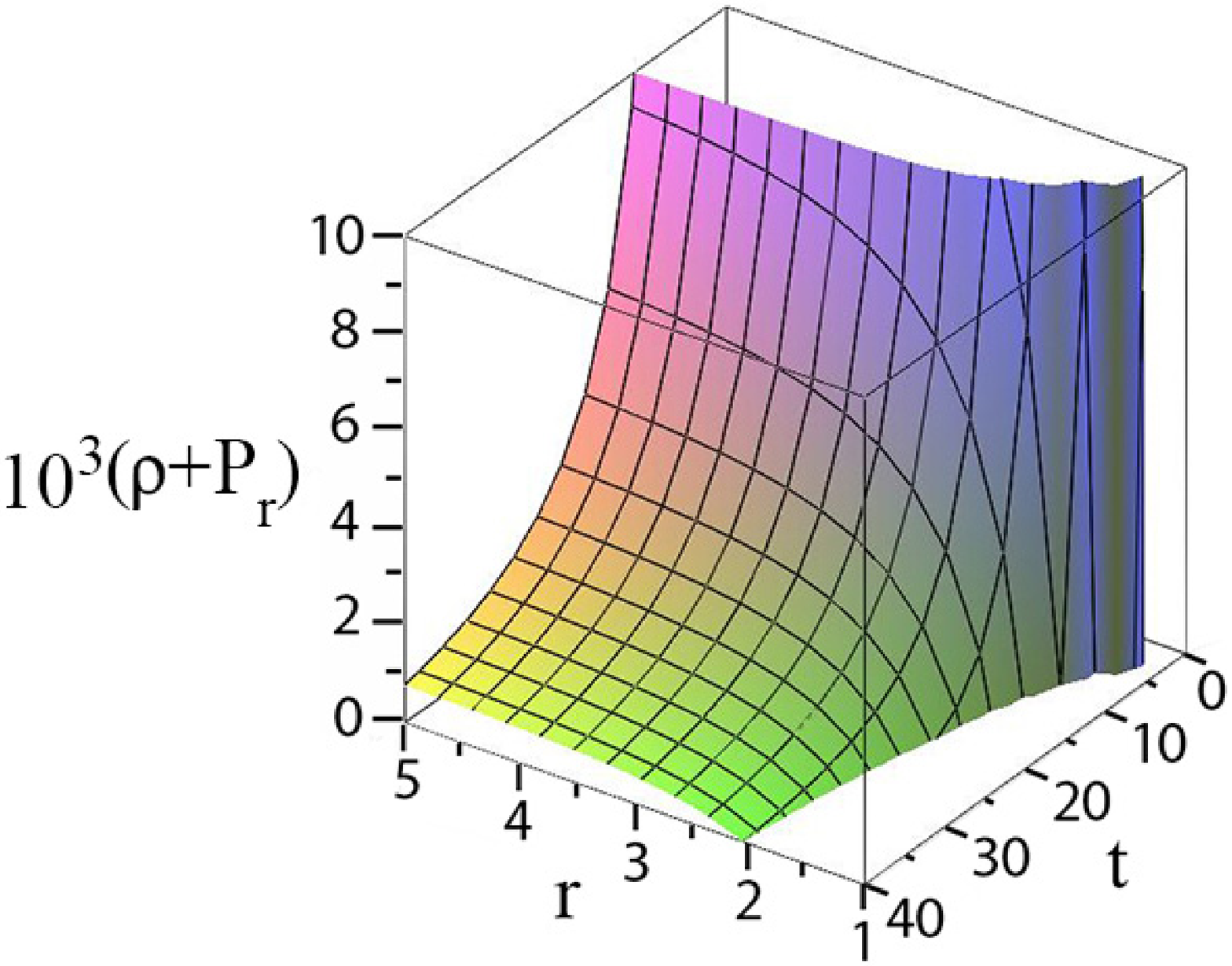} \hspace{1.5cm} %
\includegraphics[width=8cm]{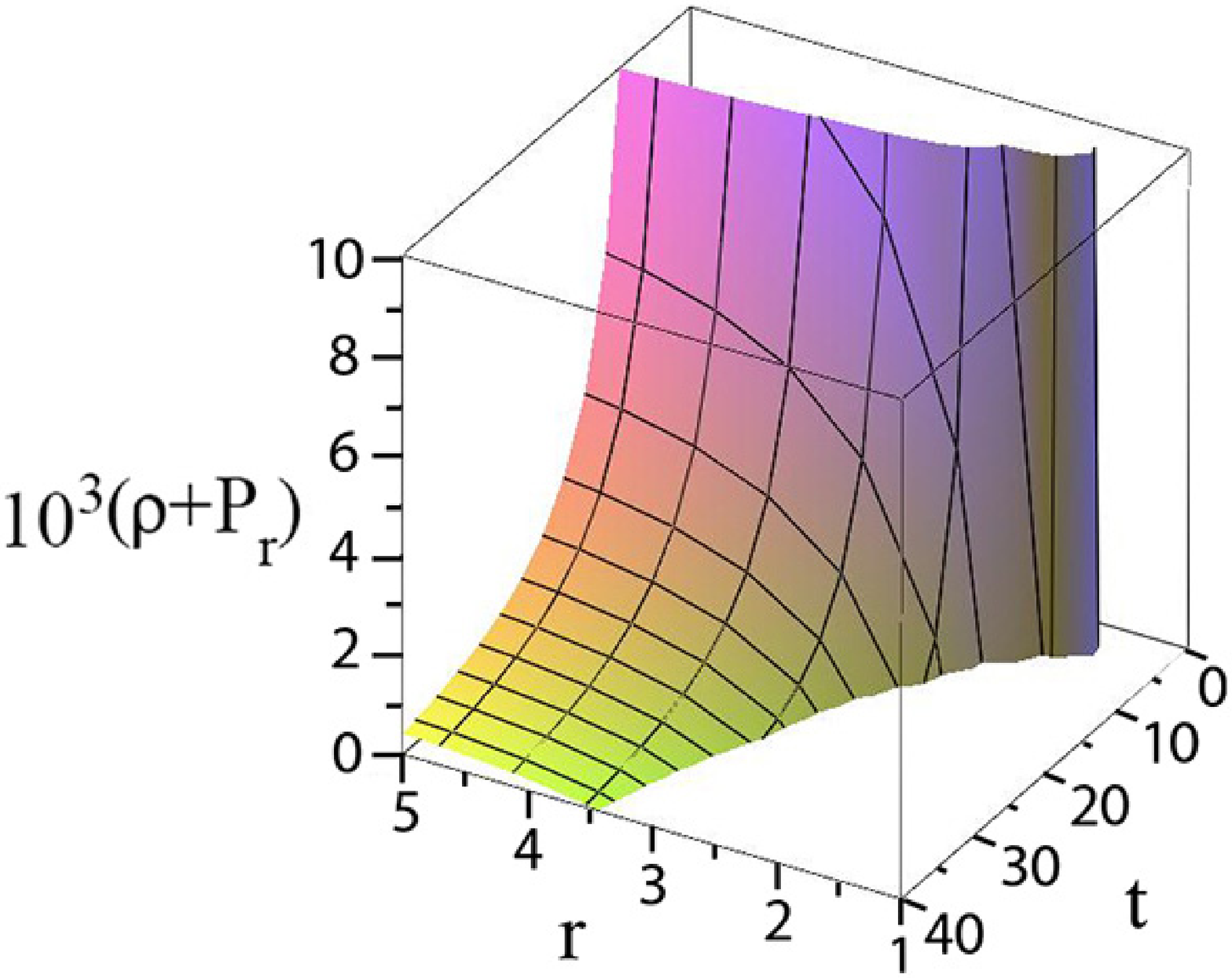}
\caption{For the flat background solution, the plots depict the behavior of
$10^3\left(\rho +P_{r}\right)$ with respect to $r$ and $t$, with $n=3$ and $A_{2}=1$,
for $\protect\omega =0$ (left plot) and $\protect\omega =1/3$ (right
plot), respectively. As it is clear from the plots, the region of the exotic
matter in the vicinity of throat increases as time passes.}
\label{i2}
\end{figure*}

We continue our discussions with the second solution for the scale factor,
for $\omega \neq -1$. In this case, we have
\begin{eqnarray}
\rho &=&{\frac{2(n-1)}{n\left( 1+\omega \right) ^{2}{t}^{2}},} \\
\rho +P_{r} &=&-\frac{\left( n-1\right) \left( n-2\right) {\mathit{r}}%
_{0}^{n-2}}{2{A}_{2}^{2}{r}^{n}}\frac{1}{{t}^{\,{\frac{4}{n\left( 1+\omega
\right) }}}}  \notag \\
&&+\frac{2\left( n-1\right) }{\left( 1+\omega \right) {n}}\frac{1}{{t}^{2}},
\\
\rho +P_{t} &=&\,\frac{\left( n-2\right) {\mathit{r}}_{0}^{n-2}}{2{A}_{2}^{2}%
{r}^{n}}\frac{1}{\,{t}^{\,{\frac{4}{n\left( 1+\omega \right) }}}}+\frac{%
2\left( n-1\right) }{\left( 1+\omega \right) {n}}\frac{1}{{t}^{2}}.
\end{eqnarray}%
For $\omega <-1$, it is clear that the NEC is violated due to $\rho +P_{r}<0$%
. For $\omega >-1$, we have that $\rho $ and $\rho +P_{t}$ are always
positive, while the quantity $\rho +P_{r}$ should be analysed in more
detail. Figure \ref{i1} shows the behaviour of $\rho +P_{r}$ with respect to
time at the throat $r_{0}=1$ for $A_{2}=1$, $n=3$ and $\omega =0$ and $1/3$.
It can be seen that although the wormhole matter at the throat initially
satisfies the NEC, the latter is violated as time passes. In Fig. \ref{i2},
the quantity $\rho +P_{r}$ is plotted against $r$ and $t$ with $r_{0}=1$, $%
A_{2}=1$, $n=3$ for $\omega=0$ and $1/3$. The figure shows that the region
of exotic matter in the vicinity of the wormhole throat increases as time
increases.

\begin{figure}[tbp]
\centering
\includegraphics[width=8cm]{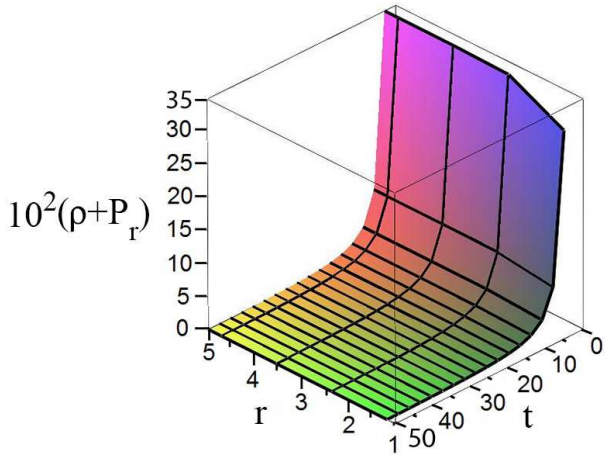}
\caption{For the flat background solution, the plot depicts the behavior of
$10^2\left(\rho +P_{r}\right)$ with respect to $r$ and $t$ for $n=4$,\ $r_{0}=1$, $%
A_{2}=2$ and $\protect\omega =-1/2$. This plot shows that for
suitable choices of constants, there are wormhole structures constructed
from matter that satisfy the null energy condition.}
\label{ii1}
\end{figure}

As we are considering higher-dimensional wormholes in an expanding
spacetime, an interesting scenario to examine is whether these dynamic
wormholes could be constructed from normal matter for $n>3$. This is indeed
the case for the solutions discussed here, by choosing suitable values for
the constants. Figure \ref{ii1} plots $\rho +P_{r}$ against $r$ and $t$ for $%
r_{0}=1$, $A_{2}=2$, $n=4$ and $\omega =-1/2$. As depicted in the figure, $%
\rho+P_{r}$ is always positive for this choice of constants and therefore
the NEC (and also WEC) is satisfied for the whole wormhole structure.

%%%%%%%%%%%%%%%%%%%%%%%%%%%%%%%%%%%%%%%%%%%%%%%%%%%%%%%%%%%%%%%%%%%%%%%%%%%%%%%%%%%%

\subsection{Open Background}

%%%%%%%%%%%%%%%%%%%%%%%%%%%%%%%%%%%%%%%%%%%%%%%%%%%%%%%%%%%%%%%%%%%%%%%%%%%%%%%%%%%%

In the case of the open background, by applying $P_{(ob)}=\omega \rho _{(ob)}
$, we consider the following analytical solutions
\begin{equation}
R(t)= \left\{
\begin{array}{l}
A_{3}\sinh (\frac{t}{A_{3}}), \qquad \text{for} \qquad \omega=-1 \\
A_{4}t, \qquad \qquad \quad \ \ \text{for} \qquad \omega = \frac{2-n}{n}%
\end{array}%
\right. ,
\end{equation}%
where the case $\omega =-1$ corresponds to the inflationary regime.

Let us first investigate the solution corresponding to the inflationary
expansion regime, $R(t)=A_{3}\sinh (t/A_{3})$. Using Eqs. (\ref{rhoopen})-(%
\ref{bgopen}), we have
\begin{eqnarray}
\rho &=&{\frac{n\left( n-1\right) }{2{A}_{3}^{2}}}, \\
\rho +P_{r}&=&-\frac{\left( n-1\right) \left( n-2\right) \left( {\mathit{r}}%
_{0}^{n}+{\mathit{r}}_{0}^{n-2}\right) }{2A_{3}^{2}{r}^{n}\sinh \left( {%
\frac{t}{A_{3}}}\right) ^{2}}\,, \\
\rho +P_{t}&=&\frac{\left( n-2\right) \left( {\mathit{r}}_{0}^{n}+{\mathit{r}%
}_{0}^{n-2}\right) }{2A_{3}^{2}{r}^{n}\sinh \left( {\frac{t}{A_{3}}}\right)
^{2}}.
\end{eqnarray}%
As in the case of the flat background, $\rho +P_{r}$ is always negative,
implying the violation of the NEC throughout the spacetime, while $\rho $
and $\rho +P_{t}$ are always positive. However, $\rho +P_{r}$ tends to zero
as $t$ increases and therefore during the inflationary era the wormhole
matter tends from an exotic matter regime to a normal matter one, at
temporal infinity.

\begin{figure}[tbp]
\centering
\includegraphics[width=8cm]{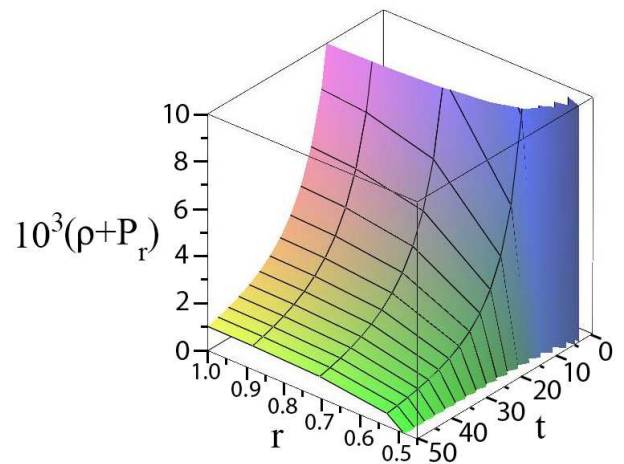}
\caption{For the open background, the plot depicts the behavior of 
$10^3\left(\rho +P_{r}\right)$ with respect to $r$ and $t$ for $n=4$, $r_{0}=0.5$ and $A_{4}=3$%
. It shows that in the case of an open universe, it is possible to have
wormholes constructed from normal matter.}
\label{ii2}
\end{figure}

Consider now the second case where $\omega =(2-n)/n$ and $R(t)=A_{4}t$. In
this case, one obtains the following relationships
\begin{eqnarray}
\rho &=&{\frac{n\left( n-1\right) \left( {A}_{4}^{2}-1\right) }{2{A}_{4}^{2}}%
}\frac{1}{{t}^{2}}, \\
\rho +P_{r} &=&\Bigg[ -\frac{\left( n-1\right) \left( n-2\right) \left( {%
\mathit{r}}_{0}^{n}+{\mathit{r}}_{0}^{n-2}\right) }{2{A}_{4}^{2}{r}^{n}}
\notag \\
&&+\frac{(n-1)\left( {A}_{4}^{2}-1\right) }{{A}_{4}^{2}}\Bigg] \frac{1}{{t}%
^{2}}, \\
\rho +P_{t} &=&\Bigg[ \frac{\left( n-2\right) \left( {\mathit{r}}_{0}^{n}+{%
\mathit{r}}_{0}^{n-2}\right) }{2{A}_{4}^{2}{r}^{n}}  \notag \\
&& +\frac{(n-1)\left( {A}_{4}^{2}-1\right) }{{A}_{4}^{2}}\Bigg] \frac{1}{{t}%
^{2}}.
\end{eqnarray}%
For $A_{4} <1$, it is clear that the NEC is violated due to $\rho +P_{r}<0$;
note also that $\rho <0$. However, this case $A_{4} <1$ should be excluded,
as $\rho $ coincides with the background energy density. This would imply
that the energy density of the universe is negative, so it is not physically
acceptable. For $A_{4} >1$, it is obvious that $\rho$ and $\rho +P_{t}$ are
always positive while $\rho +P_{r}$ should be investigated. Figure \ref{ii2}
depicts $\rho +P_{r}$ in terms of $r$ and $t$ for $r_{0}=0.5$, $A_{4}=3$ and $%
n=4$. This figure shows that by choosing suitable constants, there is a
wormhole structure constructed from normal matter.

%%%%%%%%%%%%%%%%%%%%%%%%%%%%%%%%%%%%%%%%%%%%%%%%%%%%%%%%%%%%%%%%%%%%%%%%%%%%%%%%%%%%

\section{Summary and Conclusion}

\label{conclusion}
%%%%%%%%%%%%%%%%%%%%%%%%%%%%%%%%%%%%%%%%%%%%%%%%%%%%%%%%%%%%%%%%%%%%%%%%%%%%%%%%%%%%

The present paper investigates the possibility and naturalness of expanding
wormholes in higher dimensions which is an important ingredient of
modern theories of fundamental physics, for instance, string theory, supergravity and
Kaluza-Klein, amongst others. One of our motivations for considering
wormhole solutions in an expanding cosmological background refers to the
inflationary theory where the quantum fluctuations in the inflaton field may
have served as the seed for the large scale structures in the universe. Non-trivial 
topological objects such as microscopic wormholes may have been
formed through the quantum foam and enlarged to macroscopic size during
inflation and in the subsequent expansion of the Universe. Indeed, if most
of the wormholes in the quantum foam survived enlargement through inflation,
then the Universe might be far more inhomogeneous and topologically
complicated than we observe.

Indeed, postulating higher-dimensional spacetimes is an important ingredient of modern theories of fundamental physics. In this context, the existence of higher dimensions may help construct wormhole solutions that respect energy conditions. In particular, in a cosmological set up, microscopic, dynamical wormholes produced in the early universe may be inflated to macroscopic scales and thus be -- at least in principle -- astrophysically observable. In this work, by assuming a homogeneous matter field (i.e. energy density depending only on the time coordinate), which holds in the standard cosmology,  we arrived at interestingly simple and exact solutions. More specifically, we considered a particular class of wormhole solutions corresponding to a spatially homogeneous Ricci scalar. The possibility of obtaining solutions with normal and exotic matter was explored and we found new solutions including those that satisfy the NEC in specific time intervals. In particular, in five dimensions, we found solutions that satisfy the NEC everywhere.

\acknowledgments{
FSNL acknowledges financial  support of the Funda\c{c}\~{a}o para a
Ci\^{e}ncia e Tecnologia through an Investigador FCT Research contract, with
reference IF/00859/2012, funded by FCT/MCTES (Portugal), and grants
CERN/FP/123618/2011 and EXPL/FIS-AST/1608/2013.}

\end{document}